\pdfoutput=1
\documentclass[a4paper,american,floatfix,pdftex,superscriptaddress,twoside,%
aps,pre,%
citeautoscript,%
reprint,
]{revtex4-1}%
\usepackage{amsfonts,amsmath,amssymb}
\usepackage{commath}
\usepackage[T1]{fontenc}
\usepackage{graphicx}%
\usepackage[utf8]{inputenc}
\usepackage{microtype}
\usepackage[obeyFinal,textsize=footnotesize]{todonotes}
\usepackage{xspace}
\usepackage{hyperref, hypernat}
\usepackage[todos]{CMImacros} 
\usepackage[displaymath,textmath,graphics]{preview}

\graphicspath{{./figures/}} 
\hypersetup{citebordercolor=yellow,linkbordercolor=red,urlbordercolor=blue} %

\newcommand{\ucg}{\affiliation{Department of Sciences, University College Groningen, University of
      Groningen, Hoendiepskade 23/24, 9718 BG Groningen, Netherlands}}%
\newcommand{\ucgr}{\affiliation{Groningen Biomolecular Sciences and Biotechnology Institute,
      University of Groningen, Nijenborgh 4, 9747 AG Groningen, Netherlands}}%
\newcommand{\cfeldesy}{\affiliation{Center for Free-Electron Laser Science, Deutsches
      Elektronen-Synchrotron DESY, Notkestraße 85, 22607 Hamburg, Germany}}%
\newcommand{\uhhcui}{\affiliation{Center for Ultrafast Imaging, Universität Hamburg, Luruper
      Chaussee 149, 22761 Hamburg, Germany}}%
\newcommand{\uhhphys}{\affiliation{Department of Physics, Universität Hamburg, Luruper Chaussee 149,
      22761 Hamburg, Germany}}%
\newcommand{\jkemail}{\email[Email:~]{jochen.kuepper@cfel.de}}%
\newcommand{\cmiweb}{\homepage[website:~]{https://www.controlled-molecule-imaging.org}}%

\begin{document}
\title{Microscopic force for aerosol transport}%
\author{Nils Roth}\cfeldesy\uhhphys%
\author{Muhamed Amin}\cfeldesy\ucg\ucgr%
\author{Amit K.\ Samanta}\cfeldesy%
\author{Jochen Küpper}\jkemail\cmiweb\cfeldesy\uhhphys\uhhcui
\begin{abstract}\noindent%
   A key ingredient for single particle diffractive imaging experiments is the successful and
   efficient delivery of sample. Current sample-delivery methods are based on aerosol injectors in
   which the samples are driven by fluid-dynamic forces. These are typically simulated using Stokes'
   drag forces and for micrometer-size or smaller particles, the Cunningham correction factor is
   applied. This is not only unsatisfactory, but even using a temperature dependent formulation it
   fails at cryogenic temperatures. Here we propose the use of a direct computation of the force,
   based on Epstein's formulation, that allows for high relative velocities of the particles to the
   gas and also for internal particle temperatures that differ from the gas temperature. The new
   force reproduces Stokes' drag force for conditions known to be well described by Stokes' drag.
   Furthermore, it shows excellent agreement to experiments at 4~K, confirming the improved
   descriptive power of simulations over a wide temperature range.
\end{abstract}
\maketitle

\section{Introduction}
\label{sec:introduction}
The functionality of molecules and materials is strongly correlated to their atomic structure.
Currently, biomolecules with sizes of a few nanometers are of particular interest for visualizing
their high-resolution atomic structure in order to unravel the secrets of life and for developing,
\eg, pharmaceuticals or novel biomimetic materials. With the advent of modern x-ray free-electron
lasers (XFELs) coherent-single-particle diffractive imaging (SPI) has become
feasible~\cite{Neutze:Nature406:752, Bogan:NanoLett8:310, Seibert:Nature470:78, Barty:ARPC64:415,
   Sobolev:CommPhys3:97}. SPI allows to retrieve the three-dimensional (3D) atomic structure of
nanoparticles by processing a series of two dimensional diffraction patterns of the corresponding
isolated nanoparticles \emph{in silico}.

SPI does not rely on highly-ordered crystalline sample, as in x-ray
crystallography~\cite{Shi:Cell159:995}, nor on a mechanical sample support as in cryo-electron
microscopy (CEM)~\cite{Fernandez:Nature537:339, Sugita:Nature563:137}. However, its
diffraction-before-destruction approach~\cite{Neutze:Nature406:752} requires constant replenishment
of identical targets in order to collect the necessary number of diffraction patterns for the 3D
reconstruction. Sample sources are typically aerosol injectors producing tightly focused streams of
nanoparticles~\cite{Bogan:NanoLett8:310}. However, the efficient delivery of identical nanoparticles
is still a bottleneck for SPI experiments~\cite{Bieleckie:SciAdv5:eaav8801}. Our recently reported
approach of using a cryogenic buffer-gas cooled aerosol injector~\cite{Samanta:StructDyn7:024304}
promises to overcome this limitation by increasing the reproducibility and control over the sample.
There aerosolized nanoparticles were transported into a cryogenically-cooled helium-filled
buffer-gas cell, where the nanoparticles were rapidly cooled~\cite{Samanta:StructDyn7:024304}. The
low temperature reduces particle losses and broadening of the stream due to diffusion, and it allows
for better subsequent nanoparticle control~\cite{Eckerskorn:PRAppl4:064001, Li:PRAppl11:064036,
   Chang:IRPC34:557}.

Generally, for best performance it is necessary to optimize the geometry of an aerosol injectors and
the flow conditions of the carrier gas for every individual nanoparticle sample. For SPI experiments
at room temperature simulations have already shown to be a useful tool to get insights on the sample
delivery process and to aid during optimization~\cite{Roth:JAS124:17}. However, for the cryogenic
buffer-gas cell an improved description of the interaction between the gas and the nanoparticles is
required for a better understanding of the particles' trajectories and phase-space distributions.
These simulations should also reliably predict the final temperature of the nanoparticles and their
cooling rate, an important aspect of buffer-gas cooling~\cite{Samanta:StructDyn7:024304}.

A general theory for describing the forces of an aerosol in a gas flow has yet to be found. For the
purpose of modelling particle trajectories through aerodynamic focusing devices it is important to
consider the usual working conditions that apply during the experiment. The pressure regimes can be
described by the Knudsen number $\Knud=\lambda/d_\text{P}$, which is the ratio of the mean free path
of the fluid $\lambda$ to the diameter of the particle $d_\text{P}$. In the experiment the pressure
ranges, in principle, from atmosphere to ultrahigh vacuum. However, the actual focusing and
transport that we are mainly interested in occurs in pressure regimes below 10~mbar, leading to
$\Knud\geq100$ for nanometer size particles. The regime with $\Knud\gg1$ is called molecular flow.
For this regime the boundary conditions assumed for Stokes famous drag equation do not hold any more
and an empirical correction factor to the drag force, called ``Cunningham correction factor'', was
introduced~\cite{Cunningham:PRSA83:357} and quickly improved to today's
formalism~\cite{Knudsen:AnaPhys341:981}. The empirical parameters were determined several times by
fitting the drag force to experimental data, mostly from Milikan's oil droplet
experiments~\cite{Millikan:S32:436, Millikan:PR22:1}. Hence, this original description of the drag
force is valid for the exact conditions in Milikan's experiment, namely, air at room temperature and
particles of hundreds nanometers or larger. The Cunningham correction factor depends on the gas and
temperature~\cite{Li:PRE68:061206} and the temperature-dependent correction factors were derived
from kinetic theory considerations and related to experimental data for temperatures from 200 to
1000~K~\cite{Willeke:JAS7:381}. However, we are now trying to describe experiments in helium gas at
temperatures down to 4~K~\cite{Samanta:StructDyn7:024304}.

Another approach to model the force of a rarefied fluid on a particle is to use the kinetic theory
of gases. For the momentum transfer from gas molecules impinging and emerging from the surface of a
particle, Epstein was able to reproduce the experimental data measured by Milikan by assuming 10~\%
specular reflection and 90~\% diffuse reflection and the particle to be a perfect
conductor~\cite{Epstein:PR23:710}. This approach is valid across all gas types and temperatures. For
particle sizes of current interest, which are in the order of 10--300~nm, the assumptions of a rigid
body and the mainly diffuse scattering of Epstein's model are still good approximations. For smaller
systems more advanced treatment might be necessary: For particles with sizes of a few nm specular
reflection become dominant and for small molecular sizes the treatment as rigid spheres fails and
long range interactions, \eg, van der Waals interactions and electric multipole interactions, have
to be taken into account~\cite{Li:PRE68:061206}.

The dimensions of the current aerodynamic focusing devices are on the order of millimeter to
centimeter. When miniaturizing these devices it might become necessary to include forces important
for microfluidic channels such as the Saffman force~\cite{Akhatov:JAS39:691}.

Epstein's description would match the experimental conditions~\cite{Roth:JAS124:17,
   Samanta:StructDyn7:024304} if it wasn't for the large relative velocities between particles and
gas and the temperature differences between gas and particles. These effects are incompatible with
Epstein's approach, although we note that Epstein's model was improved in several ways, \eg, to the
description of molecular-size particles based on Chapman-Enskog theory and the kinetic theory of
gases~\cite{Tammet:JAS26:459, Li:PRE68:061206}, by accounting for quantum
effects~\cite{Drosdoff:PRE71:051202}, by deriving an analytical expression for the ratio between
specular and diffuse reflection~\cite{Wiseman:CPL465:175}, through molecular dynamics
simulations~\cite{Liu:PRE99:042127, Li:PRL95:014502, Wang:ANYAS1161:484}, or for non-isothermal
fluids~\cite{Li:PRE70:021205, Wang:PRE84:021201} and lift forces due to the rotation of the particle
or the velocity gradient in the flow field~\cite{Akhatov:MN5:215, Liu:PF21:047102,
   Akhatov:JAS39:691}. Unfortunately, none of these advances treats the needed adaptation for our
experimental conditions. Hence, a new model based on Epstein's original approach is formulated.

\section{Modeling the particle transport in an aerosol injector for SPI Experiments}
\label{sec:SPI}
\subsection{Drag force in an aerosol injector for SPI experiments}
\label{sec:dragSPI}
For molecular flow the mean free path of the gas is much larger than the diameter of the particle.
Hence, it is a valid assumption that the presence of the particle does not change the gas flow, \eg,
the velocity distribution of the gas molecules. Assuming a Maxwell distribution, the number of gas
molecules with velocities between $(v_x,v_y,v_z)$ and $(v_x+\dif{v_x},v_y+\dif{v_y},v_z+\dif{v_z})$
is
\begin{multline}
   \label{eq:maxwell}
   N_{v_x,v_y,v_z}\dif{v_x}\dif{v_y}\dif{v_z} \\
   = N\left(\frac{h}{\pi}\right)^\frac{3}{2}e^{-h(v_x^2+v_y^2+v_z^2)}\dif{v_x}\dif{v_y}\dif{v_z},
\end{multline}
with
\begin{equation}
   \label{eq:beta}
   h=\frac{m}{2kT},
\end{equation}
where $N$ is the Number of molecules per unit Volume, $m$ is the mass of the gas molecule, $k$ is
the Boltzmann constant and $T$ is the gas temperature. From the point of view of a particle moving
in a gas with speed $U$ and velocity components $U_x=\alpha{U}$, $U_y=\beta{U}$ and $U_z=\gamma{U}$,
with velocities along the $x,y,x$ axes according to the fractions of speed $\alpha,\beta,\gamma$,
the velocity distribution is
\begin{multline}
   \label{eq:maxwell+U}
   N_{v_x,v_y,v_z}\dif{v_x}\dif{v_y}\dif{v_z}= \\
   N\left(\frac{h}{\pi}\right)^\frac{3}{2} e^{-h\left((v_x+\alpha U)^2+(v_y+\beta U)^2+(v_z+\gamma U)^2\right)}\dif{v_x}\dif{v_y}\dif{v_z}.
\end{multline}

To determine the amount of gas molecules that hit the particle we assumed a surface element $dS$ of
the particle normal to the $x$ direction. The volume that contains all particles with velocity
$v_x+dv_x$ that will hit the surface in unit time is given by $v_xdS$ and the amount of particles in
this volume is
\begin{equation}
   \label{eq:nop/time}
   n_{v_x,v_y,v_z}\dif{v_x}\dif{v_y}\dif{v_z}\dif{S} =v_xN_{v_x,v_y,v_z}\dif{v_x}\dif{v_y}\dif{v_z}\dif{S}.
\end{equation}

The amount of momentum transferred to the particle in a given direction by an individual gas
molecule impinging and sticking to the particle is given by $m(\alpha'v_x+\beta'v_y+\gamma'v_z)$.
For a sphere with radius $R$, the $z$-axis defined to be normal to the plane through $x$ and $U$,
and the angle $\theta$ between $y$ and $U$ we obtain
$\alpha=\cos\left(\theta\right),\beta=\sin\left(\theta\right),\gamma=0$. Furthermore, for the
momentum transferred in the direction of $U$ we obtain
$\alpha'=\cos\left(\theta\right),\beta'=\sin\left(\theta\right),\gamma=0$. The total amount of
momentum transferred in the direction of $U$ can be calculated analog to Epstein's model, directly
using \eqref{eq:maxwell+U} instead of an approximation for small $U$, by integrating over all
surface elements $dS=R^2\sin(\theta)d\theta{}d\phi$ and all gas molecules impinging the particle in
unit time. The amount of gas molecules impinging the particle per time is constant in the
statistical limit, so is the momentum transferred per time, the force. It is given by
\begin{equation}
   \label{eq:Fi}
   \begin{split}
      F_\text{imp} =& \frac{p\sqrt{\pi}R^2}{2hU^2}\left(-2e^{-hU^2}\sqrt{h}U\left(
            1+2hU^2\right)\right.  \\
      & \left. +\sqrt{\pi}\left( 1-4hU^2-4h^2U^4\right) \erf\left(\sqrt{h}U\right) \right).
   \end{split}
\end{equation}

For specular reflection the $x$ component of the velocity $U$ of all gas molecules is changing sign,
as does $\alpha$, while everything else stays the same. Performing the integration, the momentum
transferred by the reflecting gas molecules, and so the force due to reflection $F_\text{r}$,
averages to zero and the total force in case of specular reflection $F_\text{spec}$ is
\begin{equation}
   \label{eq:Fsp}
   F_\text{spec}=F_\text{imp}+F_\text{r}=F_\text{imp}.
\end{equation}

Calculating the force for diffuse scattering $F_\text{diff}$ requires to appropriately take the
temperature difference between the gas and the particle into account. Assuming the gas molecule to
thermalize to the particle's temperature during accommodation and it thus leaving the particle with
a Maxwell Boltzmann distribution according to the particles temperature~\cite{Epstein:PR23:710}, it
is possible to calculate the amount of momentum transfer by considering the conservation of the
number of gas molecules:
\begin{equation}
   \label{eq:Nequal}
   \begin{split}
   n_{v_x,v_y,v_z,\text{imp}} \, &\dif{v_x}\dif{v_y}\dif{v_z}\dif{S} = \\
   &n_{v_x,v_y,v_z,\text{leav}} \, \dif{v_x}\dif{v_y}\dif{v_z}\dif{S}
   \end{split}
\end{equation}
The left side of \eqref{eq:Nequal} is identical to \eqref{eq:nop/time} and
\begin{equation}
   \label{eq:Nleaving}
   n_{v_x,v_y,v_z, \text{leav}}=C_{\text{leav}}e^{-h'(v_x^2+v_y^2+v_z^2)}.
\end{equation}
$h'$ is defined equivalent to \eqref{eq:beta}, but using the temperature of the particle instead of
the gas temperature. Integrating \eqref{eq:Nequal} over the whole surface and all velocities,
$C_{\text{leaving}}$ is determined and thus the force on the particle:
\begin{equation}
   \label{eq:Fdiff}
   F_\text{diff}=F_\text{imp}-\frac{2}{3}\frac{h}{\sqrt{h'}}p\left(\pi\right)^\frac{3}{2}R^2U
\end{equation}
The total force is assumed to be a combination of 10~\% specular reflections and 90~\% diffuse
reflections~\cite{Epstein:PR23:710}:
\begin{equation}
   \label{eq:Ftot}
   F_\text{total}=0.1F_\text{spec}+0.9F_\text{diff}
\end{equation}

\subsection{Temperature changes of the aerosol}
\label{sec:temperatureSPI}
The drag force \eqref{eq:Fdiff} on a particle depends on its temperature. In the process of diffuse
scattering the gas molecules are assumed to thermalize to the particle's temperature. This means,
that that the velocity distribution of the impinging gas molecules differs from the the velocity
distribution of the reflected ones not only due to $U$, but also due to different temperatures.
Depending on whether the particle's temperature is higher or lower than the gas temperature, the gas
molecules take away energy from or deposit energy in the particle, respectively, in addition to the
energy deposited in kinetic energy of the particle due to $U$. We assume this additional energy
change will exclusively lead to a change in particle temperature, because it is even present with
$U=0$. Integrating over all molecules that hit the particle in unit time the change in energy is
\begin{equation}
   \label{eq:deltaE}
   \begin{split}
   \Delta E =& \frac{p\sqrt{\pi}R^2}{4hh'}\left(-2e^{-hU^2}\sqrt{h}\left(5h'+2h\left(2+h'U^2\right)\right)\right. \\
   &\left.-\frac{\sqrt{\pi}\erf\left(\sqrt{h}U\right)}{U} \left(3h'+4h^2U^2\left(2+h'U^2\right) \right.\right.\\
   &\left.\left.+4h \left(1+3h'U^2\right)\right)\right)
   \end{split}
\end{equation}
which for small values of $U$, using the same velocity approximation as in Epstein's model, simplifies to
\begin{equation}
   \label{eq:deltaE_smallU}
   \Delta E = \frac{4p\sqrt{h\pi}R^2}{h'}-\frac{4p\sqrt{\pi}R^2}{\sqrt{h}}
\end{equation}
A change of the particle's temperature is considered as a change $\Delta{E}$ of the total energy
stored in all its degrees of freedom, \ie, its specific heat $c_\text{p}$. Thus the change in
particle temperature per unit time is
\begin{equation}
   \label{eq:deltaT}
   \Delta T = \frac{\Delta E}{c_\text{p}m_\text{p}},
\end{equation}
with the particle's mass $m_\text{p}$.

\begin{figure*} 
   \includegraphics[width=\linewidth]{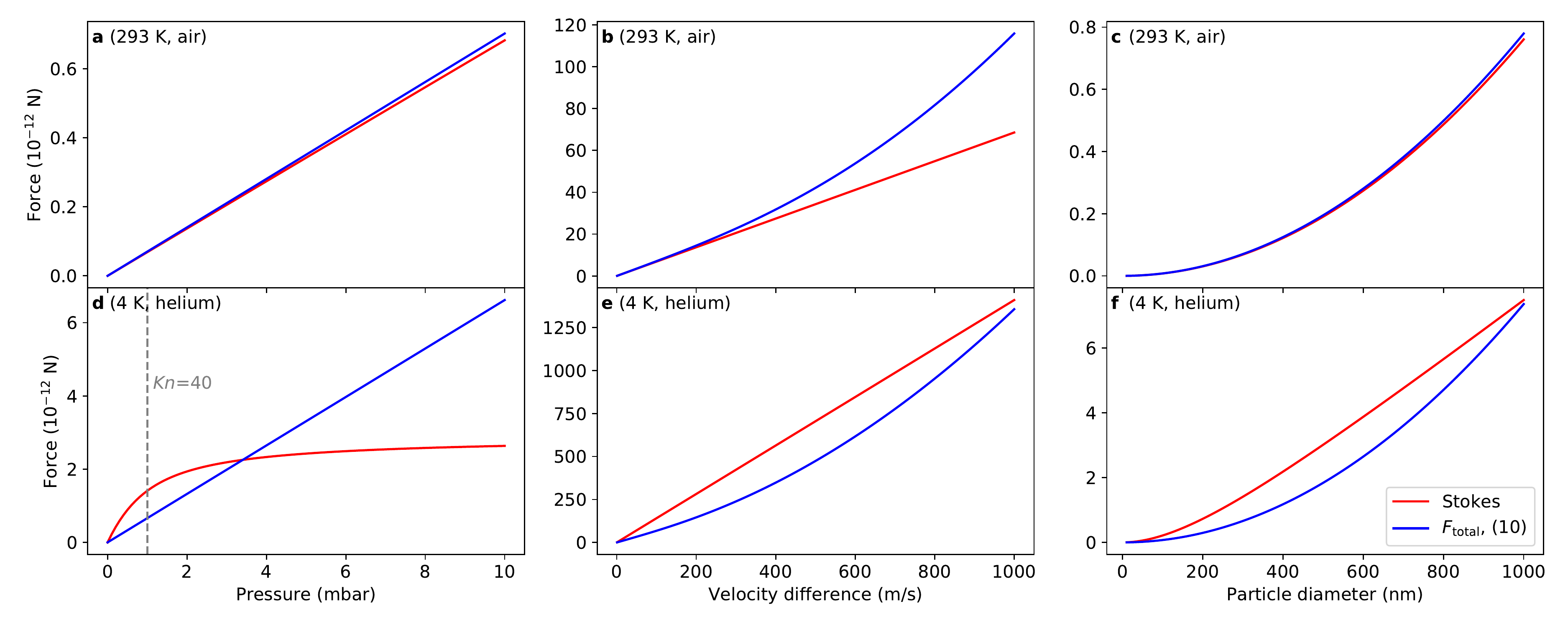}%
   \caption{Calculated drag force \eqref{eq:Ftot} in comparison to Stokes' drag force as a function
      of (a, d) pressure, (b, e) velocity difference, and (c, f) particle diameter at (a--c) room
      temperature and (d--f) 4~K. Stokes' drag force was calculated as described by
      \citet{Roth:JAS124:17} for room temperature and as described by
      \citet{Samanta:StructDyn7:024304} for 4~K. While one of the parameters is varied the others
      are fixed at 1~mbar, 1~m/s, and 300~nm, respectively. The dashed gray line in (d) indicates
      the pressure where $Kn=40$; see text for further details. }
   \label{fig:force}
\end{figure*}

\subsection{Brownian Motion}
So far we calculated the force by averaging over all single collisions the particle undergoes per
unit time, which appropriately predicts the mean force on the particle. However, its actual
trajectory depends further on its Brownian motion. For a numerical description of the Brownian
motion using the Langevin equation~\cite{Lemons:AJP65:1079} the force on the particle is split into
a part $F_\text{drag}$ that is proportional to $U$ and a part $F_b$ that is a random force.
$F_\text{drag}$ is in our case \autoref{eq:Ftot}, but using the same velocity approximation as in
Epstein's model. $F_b$ is assumed to be white noise consisting of an amplitude $A$ and a random
number $r$ with zero mean and unit variance. The fluctuation-dissipation theorem defines the
amplitude of the random force to be
\begin{equation}
   \label{eq:fluc_diss_theorem}
   A = \expectation{F_b(t_1)F_b(t_2)} = 2kT\mu\delta(t_1-t_2),
\end{equation}
with $\mu=F_\text{drag}/U$. $F_b$ considers the particle at rest with the gas and $\mu$ is
calculated for the case of small $U$, where $F_\text{total}$ is proportional to $U$. With a
numerical representation of the delta function with a time step size $\Delta{t}$ the Brownian force
is
\begin{equation}
   \label{eq:fbrown}
   F_b = r \sqrt{\frac{\left(\frac{16}{3}+\frac{2}{3}\pi\sqrt{\frac{h}{h'}}\right)
         \sqrt{\frac{\pi}{h}} p m R^2}{\Delta t}}.
\end{equation}

\section{Benchmarking the new force}
\label{sec:results}
\subsection{Comparison to Stokes' drag force}
\label{sec:comparison:stokes}
In order to validate the new force, \ie, the model derived above, we compare it to the established
model of Stokes' drag force, which is known to produce reliable results for specific conditions,
\emph{vide supra}. \autoref{fig:force} shows the calculated values of the new proposed drag force
compared to Stokes' drag force in dependence of the gas pressure, the velocity difference between
particle and gas, and the particle diameter for room temperature and 4~K, respectively.

For room temperature, \autoref[a--c]{fig:force}, and in a regime comparable to that in the Millikan
experiment both models lead to nearly identical results. When the velocity differences becomes
larger than 200~m/s the models diverge, which is expected as Stokes' drag force is only applicable
for comparable slow flows~\cite{Stokes:TCaPS9:8}, whereas \eqref{eq:Ftot} appropriately describes
that not only the amount of momentum transferred per gas molecule depends on $U$, but also the
amount of gas molecules that hit the particle increases significantly when $U$ approaches values
comparable to the average speed of a single gas molecule.

For a cold gas at 4~K, \autoref[d--f]{fig:force}, the functional behavior of the models differ.
Here, Stokes' force~\cite{Willeke:JAS7:381} is calculated as described by
\citet{Samanta:StructDyn7:024304}. For low pressures (large \Knud) both forces have a linear
pressure dependency, but with a flatter slope in case of \eqref{eq:Ftot}. In general the results
from \eqref{eq:Ftot} are below the calculated forces using Stokes in this region. However, for high
pressures (small \Knud) Stokes' force approaches a constant value. The transition occurs around
1~mbar ($\Knud\approx40$). It is important to note that the region on the left to that transition
($\Knud>>1$) is the region where the assumption for \eqref{eq:Ftot}, that the presence of the
particle is not influencing the gas flow, holds.
Smaller predicted magnitudes of the force using \eqref{eq:Ftot} can be observed in
\autoref[e,f]{fig:force} as well. These lower values are in accordance with our previous experience
using Stokes' force at these conditions: In order to successfully describe the available
experimental data using Stokes' force, it was necessary to scale the force down by roughly a factor
of 4~\cite{Samanta:StructDyn7:024304}.

\subsection{Comparison to Newton's law of cooling}
\label{sec:comparison:newton}
\begin{figure*}
   \includegraphics[width=\linewidth]{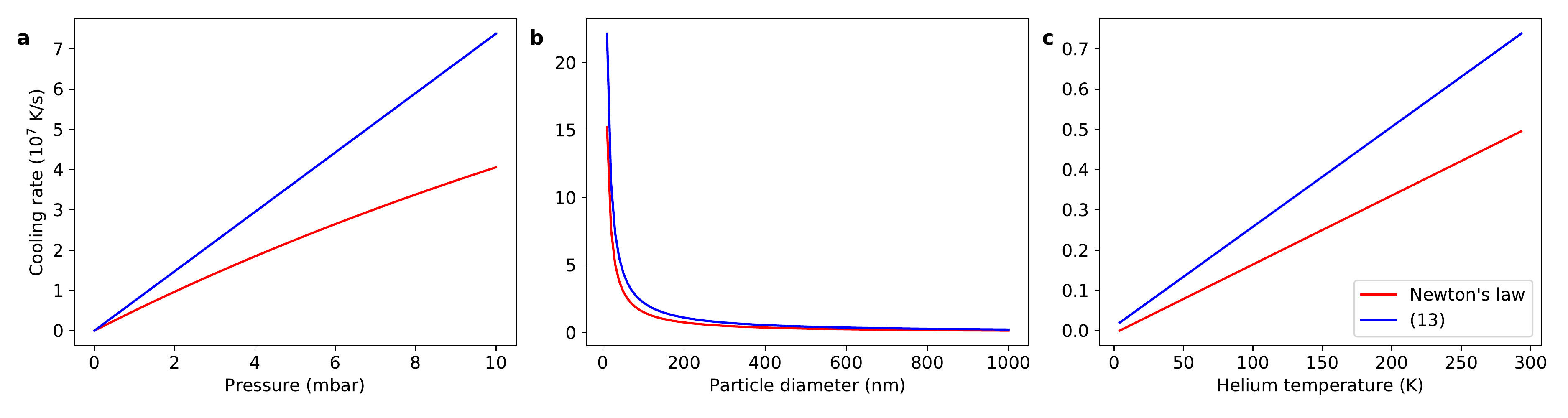}%
   \caption{Calculated values of the cooling rate \eqref{eq:deltaT} compared to Newton's law of
      cooling and its dependence on (a) pressure, (b)~particle diameter, and (c) initial temperature
      of the particle for a polystyrene sphere in helium at 4~K. While one of the parameters is
      varied the others are fixed at 1~mbar, 300~nm., 293.15~K and 0.1~m/s.}
   \label{fig:temperature}
\end{figure*}

We validated the cooling rates~\eqref{eq:deltaT}, using \eqref{eq:deltaE}, of our model against
Newton's law of cooling~\cite{Samanta:StructDyn7:024304}. The resulting cooling rates are shown in
\autoref{fig:temperature}.
\begin{figure}[b]
   \includegraphics[width=\linewidth]{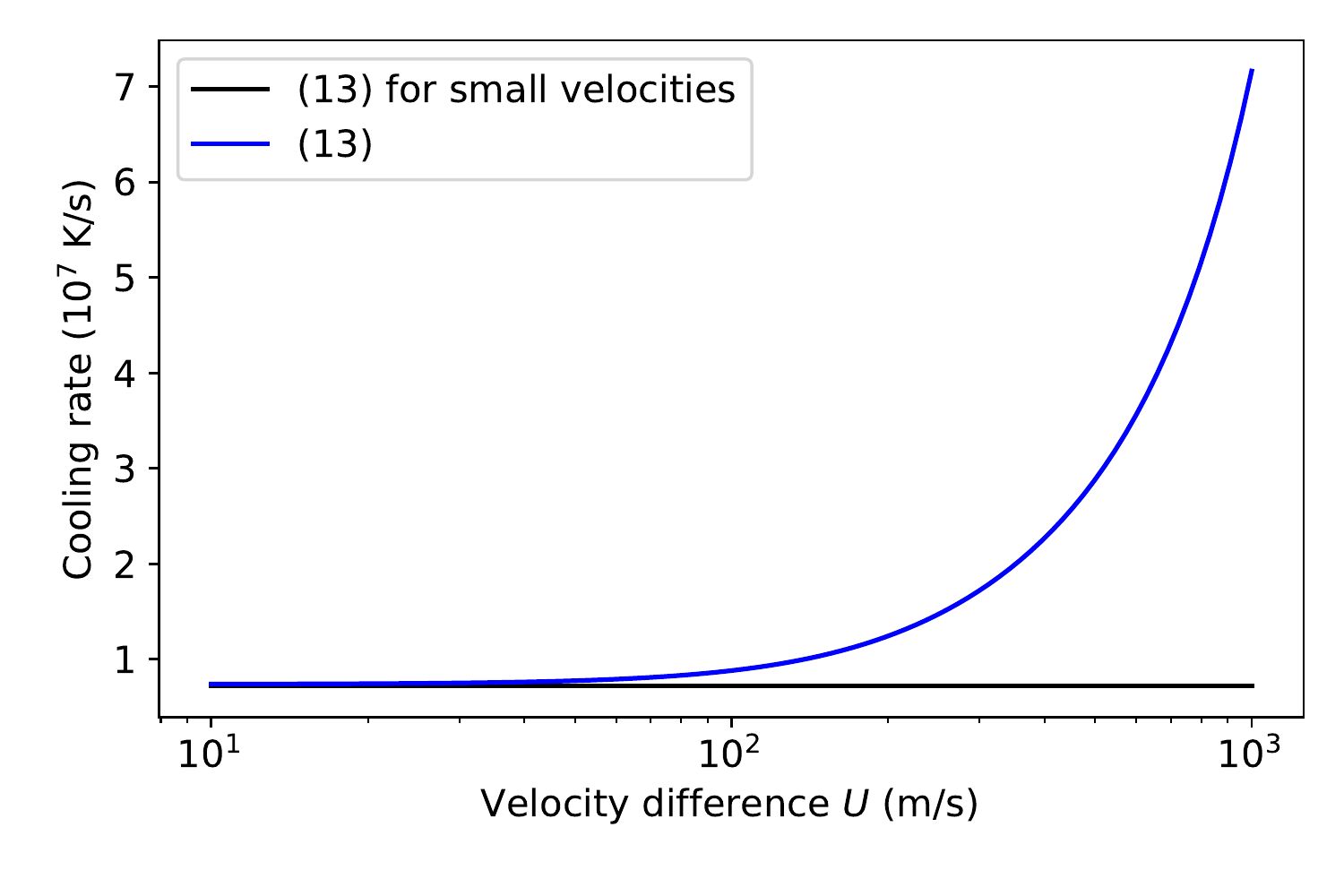}
   \caption{Cooling rate as a function of the velocity difference $U$ for a 300~nm polystyrene
      sphere in helium at 4~K and 1~mbar, calculated with (blue) the full model \eqref{eq:deltaT}
      and (black) the approximation for small velocity differences between particle and gas.}
   \label{fig:new_cooling_velocity}
\end{figure}
Newton's law of cooling and our model show the same qualitative behaviour. As expected they both
linearly depend on the temperature difference between the particle and the gas. Also the
dependencies of pressure and particle diameter are very similar. In general, our new model leads to
overall somewhat higher cooling rates, with the largest deviations roughly within a factor of two of
Newton's law of cooling.

The calculations of cooling rates using Newton's law of cooling involve several empirical
approximations in the calculation of the Nusselt number and the heat transfer coefficients for
forced convection~\cite{Samanta:StructDyn7:024304}. In addition, another empirical parameter is
needed to correct for the rarefied gas regime. The only empirical value in the new
model~\eqref{eq:deltaT} is the specific heat of the particle. Thus, it comes by no surprise, the two
models do not produce identical quantitative results and the agreement we can see in
\autoref[a--c]{fig:temperature} is pretty good, with \eqref{eq:deltaT} being a much clearer, hence
more trustworthy, model.

\autoref{fig:new_cooling_velocity} shows calculated cooling rates using the full model \eqref{eq:deltaE}
and the approximation for small velocities \eqref{eq:deltaE_smallU} for the change in energy. Up to
relative velocities of 100~m/s the calculated cooling rate does not strongly depend on velocity and
the approximation of small velocities is applicable. Hence, depending on the system of interest, it
is a valid approach to use this assumption for the sake of computational speed.

\subsection{Comparison to experimental Results}
\label{sec:exp_results}
\begin{figure}[b]
   \includegraphics[width=\linewidth]{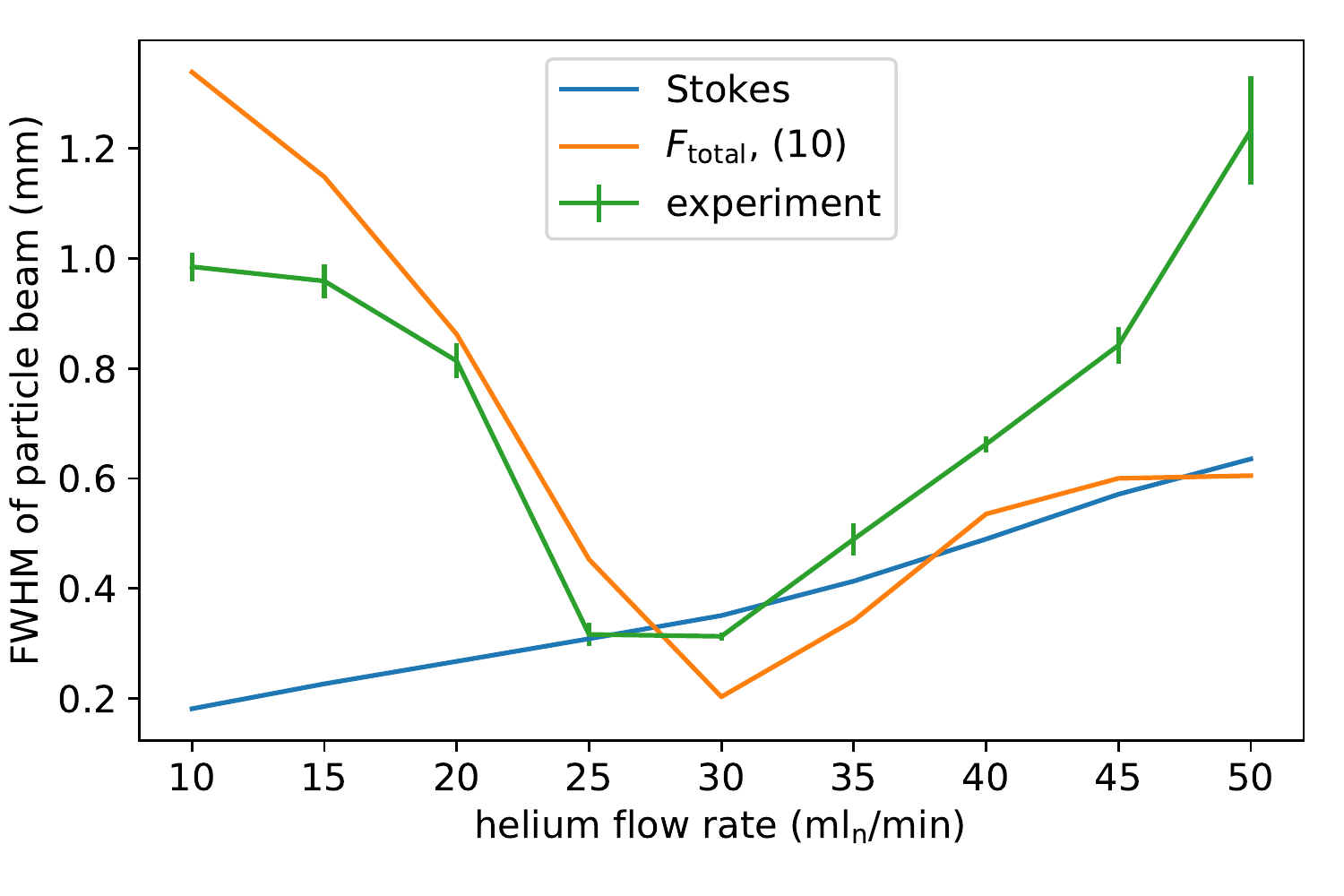}
   \caption{The full width at half maximum (FWHM) of the particle beam transverse position 1~cm
      after the buffer gas cell for different helium mass flows, simulated using the
      temperature-dependent Stokes' drag force~\cite{Willeke:JAS7:381}, simulated using
      \eqref{eq:Ftot}, and experimentally measured.}
   \label{fig:cryo}
\end{figure}
We compared our model against recent experimental and computational results for the focusing of
polystyrene spheres of diameter 220~nm in a helium buffer gas cell at
4~K~\cite{Samanta:StructDyn7:024304}. \autoref{fig:cryo} shows the full width at half
maximum~(FWHM) of the particle beams 10~mm behind the outlet of a cryogenic buffer-gas cell as a
function of the helium flow rate, \ie, differing pressures and velocities.

As expected, Stokes' drag force, even with a temperature dependent slip correction, does not
reproduce the experimental results at all, because it overestimates the force (\emph{vide supra}).
Only by scaling it down by a factor of 4 as in Fig.~4 of \cite{Samanta:StructDyn7:024304} comparable
results can be achieved.

The microscopic drag force \eqref{eq:Ftot} derived here reproduces these experimental result very
well, validating our simulation framework.

\section{Conclusion}
We have developed a new description of the flow of nanoparticles through a fluid, or the flow of a
fluid past an object, which works over a large range of pressures, relative velocities, particle
sizes, and temperatures. The model follows the ideas of Epstein's formulation of the drag force and
does not require additional empirical adjustments of the force. We have verified the model against
Stokes' drag force in the regime where the latter is valid and against experimental results for
nanoparticles at cryogenic temperatures. Our new description works very well over this wide range of
conditions.

The accurate descriptions enabled by our model are an important ingredient, for instance, for
optimized sample injection in single-particle diffraction experiments: Hit rates can be
significantly improved through reliable predictions of injection parameters before the actual
measurement campaign at the large-scale facility. This does not only improve data quality, but
allows to make much better use of the expensive x-ray pulses and thus enables better science.

As another benefit, the new model directly provides the particles' temperatures and thus the cooling
rate in the gas, which is important, for instance, for the shockfreezing of biological samples.

However, while our model is a good description for the conditions in current SPI experiments, the
envisioned advances to single-molecule samples, \ie, proteins or other macromolecules with sizes of
a few nanometers, will necessitate an advanced description of the nanoparticle-gas collisions, see
the \hyperlink{sec:introduction}{Introduction}.

The model is implemented in our larger CMInject software package for the simulation of generic
aerosol injectors, which we currently prepare for publication.

\section*{Acknowledgments}
This work has been supported by the European Research Council under the European Union's Seventh
Framework Program (FP7/2007-2013) through the Consolidator Grant COMOTION (ERC-Küpper-614507) and
the Cluster of Excellence ``Advanced Imaging of Matter'' (AIM, EXC~2056, ID~390715994) of the
Deutsche Forschungsgemeinschaft (DFG).

\bibliography{string,cmi}
\end{document}